\documentclass{article}

\usepackage{times}
\usepackage{graphicx} 
\usepackage{subfigure}

\usepackage{natbib}
\usepackage{amsmath}
\usepackage{amsfonts}
\usepackage{algorithm}
\usepackage{algorithmic}
\usepackage{multirow}
\usepackage{epstopdf}
\usepackage{bbm}
\usepackage{hyperref}

\usepackage[accepted]{icml2016}
\newtheorem{theorem}{Theorem}

\icmltitlerunning{Optimal Densification for Fast and Accurate Minwise Hashing}

\begin{document}

\twocolumn[
\icmltitle{Optimal Densification for Fast and Accurate Minwise Hashing}

\icmlauthor{Anshumali Shrivastava}{anshumali@rice.edu}
\icmladdress{Rice University, Houston, TX, 77005, USA}

\icmlkeywords{Minwise Hashing, Densified One Permutation Hashing, Large-Scale Search}

\vskip 0.3in
]

\begin{abstract}
Minwise hashing is a fundamental and one of the most successful hashing algorithm in the literature. Recent advances based on the idea of densification~\cite{Proc:OneHashLSH_ICML14,Proc:Shrivastava_UAI14} have shown that it is possible to compute $k$ minwise hashes, of a vector with $d$ nonzeros, in mere $(d + k)$ computations, a significant improvement over the classical $O(dk)$. These advances have led to an algorithmic improvement in the query complexity of traditional indexing algorithms based on minwise hashing.  Unfortunately, the variance of the current densification techniques is unnecessarily high, which leads to significantly poor accuracy compared to vanilla minwise hashing, especially when the data is sparse. In this paper, we provide a novel densification scheme which relies on carefully tailored 2-universal hashes. We show that the proposed scheme is variance-optimal, and without losing the runtime efficiency, it is significantly more accurate than existing densification techniques.  As a result, we obtain a significantly efficient hashing scheme which has the same variance and collision probability as minwise hashing. Experimental evaluations on real sparse and high-dimensional datasets validate our claims. We believe that given the significant advantages, our method will replace minwise hashing implementations in practice.\end{abstract}

\section{Introduction and Motivation}
Recent years have witnessed a dramatic increase in the dimensionality of modern datasets. \cite{Proc:Weinberger_ICML2009} show dataset with 16 trillion  ($10^{13}$) unique features. Many studies have shown that the accuracy of models keeps climbing slowly with exponential increase in dimensionality. Large dictionary based representation for images, speech, and text are quite popular~\cite{Proc:Broder,Proc:Fetterly_WWW03}. Enriching features with co-occurrence information leads to blow up in the dimensionality. 5-grams are common for text representations. With vocabulary size of $10^6$, 5-grams representation requires dimensionality of $10^{30}$. Representing genome sequences with features consisting of 32-contiguous characters (or higher)~\cite{ondov2016mash} leads to around $4^{32} = 2^{64}$ dimensions.

To deal with the overwhelming dimensionality, there is an increased emphasis on the use of hashing algorithms, such as minwise hashing. Minwise hashing provides a convenient way to obtain a compact representation of the data, without worrying about the actual dimensionality.  These compact representations are directly used in large scale data processing systems for a variety of tasks.

Minwise hashing is defined for binary vectors. Binary vectors can also be equivalently viewed as sets, over the universe of all the features, containing only attributes corresponding to the non-zero entries. Minwise hashing belongs to the {\em Locality Sensitive Hashing (LSH)} family~\cite{Proc:Broder_STOC98,Proc:Charikar}. The method applies a random permutation (or random hash function) $\pi:\Omega \rightarrow \Omega$, on the given set $S \subset \Omega$, and stores the minimum value after the permutation mapping. Formally,
 \begin{equation}h_{\pi}(S) = \min(\pi(S)).\end{equation}

Given sets $S_1$ and $S_2$, it can be shown by elementary probability arguments that
\begin{equation}
\label{eq:minhash}
Pr({h_{\pi}(S_1) = h_{\pi}(S_2)) =  \frac{|S_1 \cap S_2|}{| S_1 \cup S_2|}} = R.
\end{equation}
The quantity
\begin{equation}
 R = \frac{|S_1 \cap S_2|}{| S_1 \cup S_2|} = \frac{a}{f_1+f_2 -a},
\end{equation}
is the well known {\em Jaccard Similarity} (or resemblance) $R$ which is the most popular similarity measure in information retrieval applications~\cite{Proc:Broder}.

The probability of collision (equality of hash values), under minwise
hashing, is equal to the similarity of interest $R$. This particular
property, also known as the {\em LSH
property}~\cite{Proc:Indyk_STOC98,Proc:Charikar}, makes minwise hash
functions $h_{\pi}$ suitable for creating hash buckets, which leads to
sublinear algorithms for similarity search. Because of this same  LSH
property, minwise hashing is a popular indexing technique for a variety of
large-scale data processing applications, which include duplicate
detection~\cite{Proc:Broder,Proc:Henzinger_SIGIR06}, all-pair
similarity~\cite{Proc:Bayardo_WWW07}, temporal
correlation~\cite{Proc:Chien_WWW05}, graph
algorithms~\cite{Proc:Buehrer_WSDM08,Proc:Chierichetti_KDD09,Proc:Najork_WSDM09},
and  more.  It was recently shown that the {\em LSH property} of minwise
hashes can be used to generate kernel features for large-scale
learning~\cite{Proc:HashLearning_NIPS11}.

Minwise hashing is known to be theoretical optimal in many scenarios~\cite{DBLP:journals/corr/BavarianGHKRS16}. Furthermore, it was recently shown to be provably superior LSH for angular similarity (or cosine similarity) compared to widely popular Signed Random Projections~\cite{Proc:Shrivastava_AISTATS14}. These unique advantages make minwise hashing arguably the strongest hashing algorithm both in theory and practice.

{\bf Hashing Cost is Bottleneck:} The first step of algorithms relying on minwise hashing is to generate, some large enough, $k$ minwise hashes (or fingerprints) of the data vectors. In particular, for every data vector $x$, $h_i(x)$ $\forall i \in  \{1, 2, ..., k\}$ is repeatedly computed with independent permutations (or hash functions). These $k$ hashes are used for a variety of data mining tasks such as cheap similarity estimation, indexing for sub-linear search, kernel features for large scale learning, etc. Computing $k$ hashes of a vector $x$ with traditional minwise hashing requires $O(dk)$ computation, where $d$ is the number of non-zeros in vector $x$.  This computation of the multiple hashes requires multiple passes over the data. The number of required hashes typically ranges from few hundreds to several thousand. For example, the number of hashes required by the famous LSH algorithm is $O(n^\rho)$ which grows with the size of the data. ~\cite{Proc:Li_KDD15} showed the necessity of around 4000 hashes per data vector in large-scale learning. Hashing time is the main computational and resource bottleneck step in almost all applications using minwise hashing.

{\bf Other Related Fast Sketches are not LSH:} Two notable techniques for estimating Jaccard Similarity are: 1) bottom-$k$ sketches and 2) one permutation hashing~\cite{Proc:Li_Owen_Zhang_NIPS12}. Although these two sketches are cheap to compute, they do not satisfy the key {\em LSH property} and therefore are unsuitable for replacing minwise hashing~\cite{Proc:OneHashLSH_ICML14,Proc:Shrivastava_UAI14}. There are also substantial empirical evidence that using these (non-LSH) sketches for indexing leads to a drastic bias in the expected behavior, leading to poor accuracy.

{\bf The Idea of ``Densified" One Permutation Hashing:} Recently, ~\cite{Proc:OneHashLSH_ICML14} showed a technique for densifying sparse sketches from one permutation hashing which provably removes the bias associated with one permutation hashing. Please see Section~\ref{sec:densifi} for details. This was the first success in creating efficient hashing scheme which satisfies the {\em LSH property} analogous to minwise hashing and at the same time the complete process only requires $O(d+k)$ computations instead of the traditional bottleneck of $O(dk)$. Having such an efficient scheme directly translates into algorithmic improvements for a variety of machine learning and data mining tasks.

{\bf Current Densification is Inaccurate For Very Sparse Datasets:} The densification process although efficient and unbiased was shown to have unnecessarily higher variance. It was shown in~\cite{Proc:Shrivastava_UAI14} that the traditional ``densification" lacks sufficient randomness. It was further revealed that densification could be provably improved by using $k$ extra random bits. The improved scheme has reduced variance, and it retained the computational efficiency, see~\cite{Proc:Shrivastava_UAI14} for more details. An improved variance was associated with a significant performance gain in the task of near-neighbor search. In this work, we show that even the improved densification scheme is far from optimal.  The findings of~\cite{Proc:Shrivastava_UAI14} leaves an open curiosity: What is the best variance that can be achieved with ``densification" without sacrificing the running time? We close this by providing a variance-optimal scheme.

{\bf Our Contributions:} We show that the existing densification schemes, for fast minwise hashing, are not only sub-optimal but, worse, their variances do not go to zero with increasing number of hashes. The variance with an increase in the number of hashes converges to a positive constant. This behavior implies that increasing the number of hashes after a point will lead to no improvement, which is against the popular belief that accuracy of randomized algorithms keeps improving with an increase in the number of hashes.

To circumvent these issues we present a novel densification scheme which has provably superior variance compared to existing schemes. We show that our proposal has the optimal variance that can be achieved by densification. Furthermore, the variance of new methodology converges to zero with an increase in the number of hashes, a desirable behavior absent in prior works.  Our proposal makes novel use of 2-universal hashing which could be of independent interest in itself. The benefits of improved accuracy come with no loss in computational requirements, and our scheme retains the running time efficiency of the densification.

We provide rigorous experimental evaluations of existing solutions concerning both accuracy and running time efficiency, on real high-dimensional datasets. Our experiments validate all our theoretical claims and show significant improvement in accuracy, comparable to minwise hashing, with a significant gain in computational efficiency.

\section{Important Notations and Concepts}

Equation~\ref{eq:minhash} (i.e. the {\em LSH Property}) leads to an estimator of Jacard Similarity $R$, using $k$ hashes, defined by:
\begin{equation}\label{eq:est}
   \hat{R} = \frac{1}{k}\sum_{i=1}^k {\bf 1}(h_i(S_1) = h_i(S_2)).
\end{equation}
Here ${\bf 1}$ is the indicator function. In the paper, by variance, we mean the variance of the above estimator. Notations like $Var(h^+)$, will mean the variance of the above estimator when the $h^+$ is used as the hash function.

$[k]$ will denote the set of integers $\{1,2,...,k\}$. $n$ denotes the number
of points (samples) in the dataset. $D$ will be used for dimensionality. We
will use $\min\{S\}$ to denote the minimum element of the set $S$. A
permutation $\pi: \Omega \rightarrow \Omega$ applied to a set $S$ is another set $\pi(S)$, where $x \in S$ if and only if
$\pi(x) \in \pi(S)$. Our hashing will generate $k$ hashes $h_i \ i \in\{1,2,...,k\}$, generally from different bins. Since they all have same distribution and properties we will drop subscripts. We will use $h$ and $h^+$ to denote the hashing schemes of~\cite{Proc:OneHashLSH_ICML14} and ~\cite{Proc:Shrivastava_UAI14} respectively.

\section{Background: Fast Minwise Hashing via Densification}
\subsection{2-Universal Hashing}
\label{sec:2univ}
{\bf Definitions:} A randomized function $h_{univ}: [l] \rightarrow [k]$ is 2-universal if for all, $i, j \in [l]$ with $i \ne j$, we have the following property for any $z_1, z_2 \in [k]$
\begin{align}
  Pr(h_{univ}(i) = z_1 \mbox{ and } h_{univ}(j) = z_2 ) = \frac{1}{k^2}
\end{align}

~\cite{Proc:Carter_STOC77} showed that the simplest way to create a 2-universal hashing scheme is to pick a prime number $p \ge k$, sample two random numbers $a, \ b$ and compute
$$h_{univ}(x) = ((ax+b)~{\bmod  ~}p)~{\bmod  ~}k$$

\subsection{One Permutation Hashing and Empty Bins}

It was shown in ~\cite{Proc:Li_Owen_Zhang_NIPS12,dahlgaard2015hashing} that instead of computing the global minimum in Equation~\ref{eq:minhash}, i.e., $h(S) = \min(\pi(S))$, an efficient way to generate $k$ sketches, using one permutation,  is to first bin the range space of $\pi$, i.e. $\Omega$, into $k$ disjoint and equal partitions followed by computing minimum in each bin (or partition).

Let $\Omega_i$ denote the $i^{th}$ partition of the range space of $\pi$, i.e. $\Omega$. Formally, the $i^{th}$ one permutation hashes (OPH) of a set $S$ is defined as
\begin{equation}\label{eq:OPH}
  h^{OPH}_i(S) =  \begin{cases}
                    \min \{\pi(S) \cap \Omega_i \}, & \mbox{if } \{\pi(S) \cap \Omega_i\} \ne \phi \\
                    E, & \mbox{otherwise}.
                  \end{cases}
\end{equation}

\begin{figure}[t]
\vspace{-0.5in}
\begin{center}
\mbox{
\hspace{-0.5in}\includegraphics[width=4.3in]{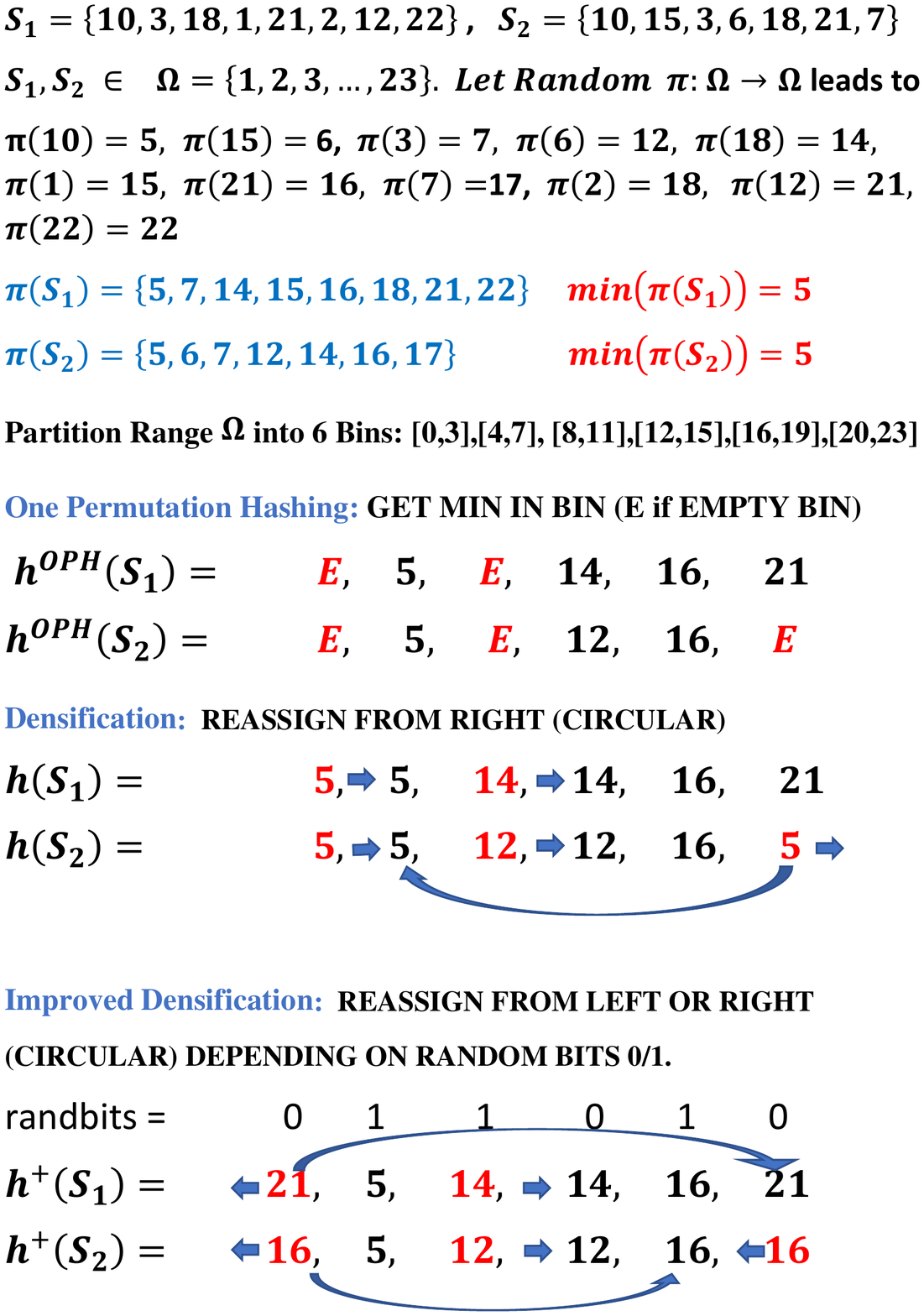}}
\end{center}
\vspace{-0.4in}
\caption{Illustration of One Permutaion Hashing (OPH) and the two existing Densification Schemes of~\cite{Proc:OneHashLSH_ICML14,Proc:Shrivastava_UAI14}. Densification simply borrows the value from near-by empty bins. Different sets $S_i$ with have different pattern of empty/nonempty bins. Since the pattern is random, the  process is similar to random re-use of unbiased values, which satisfies LSH property in each bin~\cite{Proc:Shrivastava_UAI14}.}
\label{fig:Back}
\vspace{-0.15in}
\end{figure}

An obvious computational advantage of this scheme is that it is likely to generate many hash values, at most $k$, and only requires one permutation $\pi$ and only pass over the sets (or binary vectors) $S$.
It was shown that for any two sets $S_1$ and $S_2$ we have a conditional collision probability similar to minwise hashing.
\begin{align}\label{eq:semp}
  \mbox{Let \ \ \ } E_i = \mathbf{1}\big\{ h^{OPH}_i(S_2) = h^{OPH}_i(S_2) =E \big\} \\\label{eq:OPH_CP}
  Pr\big(h^{OPH}_i(S_1) = h^{OPH}_i(S_2) \ \big| \ E_i = 0\big) = R\\
  \mbox{However, } Pr\big(h^{OPH}_i(S_1) = h^{OPH}_i(S_2) \ \big| \ E_i = 1\big) \ne R
\end{align}
Here $E_i$ is an indicator random variable of the event that the $i^{th}$ partition corresponding to both $S_1$ and $S_2$ are empty. See Figure~\ref{fig:Back}

Any bin has a constant chance of being empty. Thus, there is a positive probability of the event $\{E_i =1\}$, for any given pair $S_1$ and $S_2$ and hence for large datasets (big $n$) a constant fraction of data will consist of simultaneously empty bins (there are $n^2 \times k$ trials for the bad event $\{E_i =1\}$ to happen). This fraction further increases significantly with the sparsity of the data and $k$, as both sparsity and $k$ increases the probability of the bad event $\{E_i =1\}$. See Table~\ref{tab:statebin} for statistics of empty bins on real scenarios.

Unfortunately, whenever the outcome of the random permutation leads to simultaneous empty bins, i.e. event $E_i = 1$, the {\em LSH Property} is not valid. In fact, there is not sufficient information present in the simultaneous empty partitions for any meaningful statistics. Hence, one permutation hashing cannot be used as an LSH. Simple heuristics of handling empty bins as suggested in~\cite{Proc:OneHashLSH_ICML14} leads to a significant bias and it was shown both theoretically and empirically that this bias leads to significant deviation from the expected behavior of one permutation hashing when compared with minwise hashing. Thus, one permutation hashing although computationally lucrative is not a suitable replacement for minwise hashing.

\subsection{The Idea of Densification}
\label{sec:densifi}
In~\cite{Proc:OneHashLSH_ICML14}, the authors proposed ``densification" or reassignment of values to empty bins by reusing the information in the non-empty bins to fix the bias of one permutation hashing.  The overall procedure is quite simple. Any empty bin borrows the values of the closest non-empty bins towards the circular right (or left)\footnote{In~\cite{Proc:OneHashLSH_ICML14} they also needed an offset because the value of a hash in any bin was always reset between [0,k]. We do not need the offset if we use the actual values of $\pi(S)$. }. See Figure~\ref{fig:Back} for an illustration. Since the positions of empty and non-empty bins were random, it was shown that densification (or reassignment) was equivalent to a stochastic reselection of one hash from a set of existing informative (coming from non-empty bins) hashes which have the LSH property. This kind of reassignment restores the {\em LSH property} and collision probability for any two hashes, after reassignment, is exactly same as that of minwise hashing.

The densification generates $k$ hashes with the required {\em LSH Property} and only requires two passes over the one permutation sketches making the total cost of one permutation hashing plus densification $O(d+k)$. This was a significant improvement over $O(dk)$ with classical minwise hashing. $O(d+k)$ led to an algorithmic improvement over randomized algorithms relying on minwise hashing, as hash computation cost is bottleneck step in all of them.

\subsection{Lack of Randomness in Densification}

It was pointed out in~\cite{Proc:Shrivastava_UAI14} that the densification scheme of~\cite{Proc:OneHashLSH_ICML14} has unnecessarily high variance. In particular, the probability of two empty bins borrowing the information of the same non-empty bin was significantly higher. This probability was due to poor randomization (load balancing) which hurts the variance. ~\cite{Proc:Shrivastava_UAI14} showed that infusing more randomness in the reassignment process by utilizing $k$ extra random bits provably improves the variance. See Figure~\ref{fig:Back} for an example illustration of the method.  The running time of the improved scheme was again $O(d+k)$ for computing $k$ hashes. This improvement retains the required {\em LSH property}, however this time with improved variance. An improved variance led to significant savings in the task of near-neighbor search on real sparse datasets.

\section{Issues with Current Densification}

Our careful analysis reveals that the variance, even with the improved scheme, is still significantly higher. Worse, even in the extreme case when we take $k \rightarrow \infty$ the variance converges to a positive constant rather than zero, which implies that even with infinite samples, the variance will not be zero. This positive limit further increases with the sparsity of the dataset. In particular, we have the following theorem about the limiting variances of existing techniques:

\begin{theorem}
\label{theo:VarConst}
  Give any two finite sets $S_1, S_2 \in \Omega$, with $A = |S_1 \cup S_2|  > a = |S_1 \cap S_2| > 0$ and $|\Omega| = D \rightarrow \infty$. The limiting variance of the estimators from densification and improved densification when $k =D \rightarrow \infty$ is given by:
  \begin{small}
  \begin{align}\label{eq:limit1}
    \lim_{k \rightarrow \infty} Var(h) =  \frac{a}{A}\bigg[\frac{A -a}{A(A+1)}\bigg] &> 0  \\\label{eq:limit2}
    \lim_{k \rightarrow \infty} Var(h^+) =  \frac{a}{A}\bigg[\frac{3(A-1) + (2A-1)(a-1)}{2(A+1)(A-1)}& - \frac{a}{A}\bigg]> 0
  \end{align}
  \end{small}
\end{theorem}
This convergence of variance to a constant value, despite infinite samples, of the existing densification is also evident in our experimental findings (see Figure~\ref{fig:SimEst}) where we observe that the MSE (Mean Square Error) curves go flat with increasing $k$. Similar phenomenon was also reported in~\cite{Proc:Shrivastava_UAI14}. It should be noted  that for classical minwise hashing the variance is $\frac{R(1-R)}{k} \rightarrow 0$ for any pair $S_1$ and $S_2$. Thus, current densification, although fast, loses significantly in terms of accuracy. We remove this issue with densification. In particular, we show in Theorem~\ref{theo:optDens} that the limiting variance of the proposed optimal densification goes to 0. Our experimental findings suggests that the new variance is very close to the classical minwise hashing.  In addition, the new densification retains the speed of existing densified hashing thereby achieving the best of the both worlds.
\section{Optimal Densification}

We argue that even with the improved densification there is not enough randomness (or load balancing) in the re-assignment process which leads to reduced variance.

For given set $S$, the densification process reassigns every empty bin with a value from one of the existing non-empty bins. Note, the identity of empty and non-empty bins are different for different sets. To ensure the LSH property, the re-assignment should be consistent for any given pair of sets $S_1$ and $S_2$. In particular, as noted in~\cite{Proc:OneHashLSH_ICML14}, given any arbitrary pair $S_1$ and $S_2$, whenever any given bin $i$ is simultaneously empty, i.e. $E_i =1$, the reassignment of this bin $i$ should mimic the collision probability of one of the simultaneously non-empty bin $j$ with $E_j =0$. An arbitrary reassignment (or borrow) of values will not ensure this consistency across all pairs. We would like to point out that the reassignment of $S_1$ has no idea about $S_2$ or any other object in the dataset. Thus, ensuring the consistency is non-trivial. Although the current densification schemes achieve this consistency by selecting the nearest non-empty bin (as shown in~\cite{Proc:Shrivastava_UAI14}), they lack sufficient randomness.

\subsection{Intuition: Load Balancing}

\begin{figure}[t]
\vspace{-0.6in}
\begin{center}
\mbox{
\hspace{-0.5in}\includegraphics[width=4.3in]{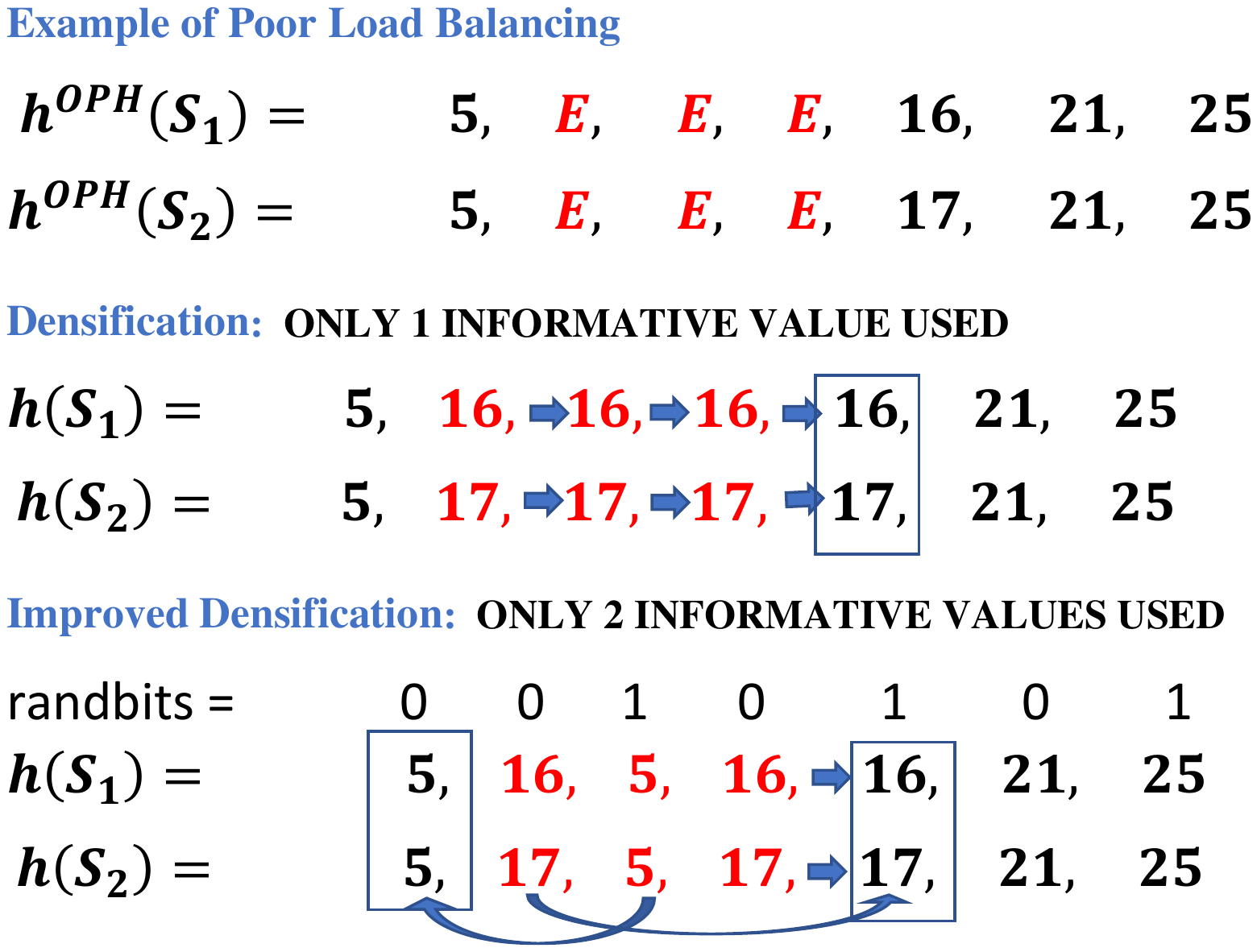}}
\end{center}
\vspace{-2.8in}
\caption{Illustration of Poor Load Balancing in the existing Densification strategies. The re-assignment of bins is very local and its not uniformly distributed. Thus, we do not use the information in far off non-empty bins while densification.}
\label{fig:PoorLoadB}
\vspace{-0.2in}
\end{figure}

In Figure~\ref{fig:PoorLoadB}, observe that if there are many contiguous non-empty bins (Bins 2, 3 and 4), then with densification schemes $h$, all of them are forced to borrow values from the same non-empty bin (Bin 5 in the example). Even though there are other informative bins (Bins 1, 6 and 7), their information is never used.  This local bias increases the probability ($p$) that two empty bins get tied to the same information, even if there are many other informative non-empty bins. Adding $k$ random bits improves this to some extent by allowing load sharing between the two ends instead of one (Bins 1 and 5 instead of just 5). However, the load balancing is far from optimal. The locality of information sharing is the main culprit with current densification schemes. Note, the poor load balancing does not change the expectation but affects the variance significantly.

For any given pairs of vectors $S_1$ and $S_2$, let $m$ be the number of simultaneous non-empty bins (out of $k$), i.e. $\sum_{i=1}^{k}E_i = k-m$. Note, $m$ is a random variable whose value is different for every pair and depends on the outcome of random $\pi$. Formally, the variance analysis of~\cite{Proc:Shrivastava_UAI14} reveals that the probability that any two simultaneous empty bin $p$ and $q$ ($E_p = E_q= 1$) reuses the same information is $\frac{2}{m+1}$ with the densification scheme $h$. This probability was reduced down to $\frac{1.5}{m+1}$ with $h^+$ by utilizing $k$ extra random bits  to promote load balancing.

$p = \frac{1.5}{m+1}$ is not quite perfect load balancing. In a perfect load balancing with $m$ simultaneous non-empty bins, the probability of two empty bins hitting the same non-empty bins is at best $p =\frac{1}{m}$. Can we design a densification scheme which achieves this $p$ while maintaining the consistency of densification and at the same time does not hurt the running time? It is not clear if such a scheme even exists. We answer this question positively by constructing a densification method with precisely all the above requirements. Furthermore we show that achieving $p = \frac{1}{m}$ is sufficient for having the limiting variance of zero.

\subsection{Simple 2-Universal Hashing Doesn't Help}

To break the locality of the information reuse and allow a non-empty bin to borrow information from any other far off bin consistently, it seems natural to use universal hashing. The hope is to have a 2-universal hash function (Section~\ref{sec:2univ}) $h_{univ}: [k] \rightarrow [k]$. Whenever a bin $i$ is empty, instead of borrowing information from neighbors, borrow information from bin $h_{univ}(i)$. The hash function allows consistency across any two $S_1$ and $S_2$ hence preserves {\em LSH property}.  The value of $h_{univ}(i)$ is uniformly distributed, so any bin is equally likely. Thus, it seems to break the locality on the first thought. If $h_{univ}(i)$ is also empty then we continue using $h_{univ}(h_{univ}(i))$ until we reach a non-empty bins whose value we re-use.

One issue is that of cycles. If $i = h_{univ}(i)$ (which has $\frac{1}{k}$ chance), then this creates a cycle and the assignment will go into infinite loop. A cycle can even occur if $h_{univ}(h_{univ}(i)) = i$ with both $i$ and $h_{univ}(i)$ being empty. Note, the process runs until it finds a non-empty bin. However, cycles are not just our concern. Even if we manage to get away with cycles, this scheme does not provide the required load balancing.

A careful inspection reveals that there is a very significant chance that both $i$ and $h_{univ}(i)$ to be empty for any given set $S$. Observe that if $i$ and $h_{univ}(i)$ are both empty, then we are bound to reuse the information of the same non-empty bin for both empty bins $i$ and $h_{univ}(i)$. We should note that we have no control over the positions of empty and non-empty bins. In fact, if no cycles happen then it is not difficult to show that the simple assignment using universal hashing is equivalent to the original densification $h$ with the order of bins reshuffled using $h_{univ}(.)$. It has worse variance than $h^+$.

\subsection{The Fix: Carefully Tailored 2-Universal Hashing}

\begin{algorithm}[ht]
\begin{algorithmic}
\INPUT $k$ One Permutation Hashes $h^{OPH}[ \ ]$ of $S$.
\INPUT $h_{univ}(.,.)$

\vspace{0.1in}

\STATE Initialize $h^*[ \ ] = 0$
\FOR{$i = 1 \ to \ k$}
\IF{$OPH[i] \not= E$}
\STATE $h^*[i] =  h^{OPH}[i]$
\ELSE
\STATE $attempt = 1$
\STATE $next = h_{univ}(i,attempt)$
\WHILE{$ h^{OPH}[next] \not= E$}
\STATE $attempt++$
\STATE $next = h_{univ}(i,attempt)$
\ENDWHILE
\STATE $h^*[i] =  h^{OPH}[next]$
\ENDIF
\ENDFOR

\STATE RETURN $h^*[ \ ]$
\end{algorithmic}
\caption{Optimal Densification}
\label{alg:optdens}
\end{algorithm}

It turns out that there is a way to use universal hashing that ensures no cycles as well as optimal load balancing. We describe the complete process in Algorithm~\ref{alg:optdens}. The key is to use a 2-universal hashing $h_{univ}: [k] \times \mathbb{N} \rightarrow [k]$ which takes two arguments: 1) The current bin id that needs to be reassigned and 2) the number of failed attempt made so far to reach a non-empty bin. This second argument ensures no infinite loops as it changes with every attempt. So even if we reach the same non-empty bin back (cycle), the next time we will visit a new set of bins. Also, even if both $i$ and $j = h_{univ}(i,attempt_i)$ are empty, $i$ and $j$ are not bound to end to the same non-empty bin. This is because in the next attempt we seek bin value $h_{univ}(i, attempt_i+1)$ for $i$ which is independent of the $h_{univ}(j,attempt_j)$ due to 2-universality of the hash function $h_{univ}$. Thus, the probability that any two empty bins reuse the information of the same non-empty bin is $\frac{1}{m}$

\subsection{Analysis and Optimality}

We denote the final $k$ hashes generated by the proposed densification scheme of Algorithm~\ref{alg:optdens} using $h^*$ (* for optimality). Formally, with the optimal densification $h^*$, we have the following:
\begin{theorem}
\label{theo:optDens}
  \begin{align}\label{eq:optDens}
    Pr\big(h^*(S_1) &= h^*(S_2)\big) = \frac{|S_1 \cap S_2|}{|S_1 \cap S_2|}= R \\\label{eq:thevar}
    Var(h^*) &= \frac{R}{k} + A \frac{R}{k^2} + B  \frac{R\bar{R}}{k^2} - R^2 \\
    \lim_{k \rightarrow \infty} Var(h^*) &= 0
  \end{align}
  where $N_{emp}$ is the number of simultaneous empty bins between $S_1$ and $S_2$ and the quantities $A$ and $B$ are given by
  \begin{small}
  \begin{align}\notag
    A &= \mathbb{E}\bigg[2N_{emp} + \frac{N_{emp}(N_{emp}-1)}{k-N_{emp}}\bigg] \\\notag
    B &=  \mathbb{E}\bigg[(k-N_{emp})(k - N_{emp}-1) + 2N_{emp}(k-N_{emp}-1)\\\notag &+ \frac{N_{emp}(N_{emp}-1)(k-N_{emp}-1)}{k-N_{emp}}\bigg]
  \end{align}
  \end{small}
\end{theorem}

Using the formula for $Pr(N_{emp} = i)$ from~\cite{Proc:Li_Owen_Zhang_NIPS12}, we can precisely compute the theoretical variance. The interesting part is that we can formally show that the variance of the proposed scheme is strictly superior compared to the densification scheme with random bits improvements.
\begin{theorem}
  \begin{align}\label{eq:varcomp}
      Var(h^*) \le Var(h^+) \le Var(h)
  \end{align}
\end{theorem}

Finally, due to optimal pairwise load balancing the variance is the best possible we can hope with independent reassignments. Formally, we have
\begin{theorem}
   Among all densification schemes, where the reassignment process for bin $i$ is independent of the reassignment process of any other bin $j$, Algorithm~\ref{alg:optdens} achieves the best possible variance.
\end{theorem}

{\bf Note:} The variance can be reduced if we allow correlations in the assignment process, for example if we force bin $i$ and bin $j$ to not pick the same bin during reassignments, this will reduce $p$ beyond the perfectly random load balancing value of $\frac{1}{m}$. However, such tied reassignment will require more memory and computations for generating structured hash functions.

\subsection{Running Time}

\begin{table}[t]
\centering
\begin{tabular}{|c|c|c|c|} \hline
    & Sim $R$ & $|S_1|$ & $|S_2|$ \\\hline
Pair 1& 0.94 & 208 & 196 \\\hline
Pair 2& 0.87 & 47 & 41 \\\hline
Pair 3& 0.74 & 91 & 67 \\\hline
Pair 4&0.61 & 15 & 14 \\\hline
Pair 5& 0.54 & 419 & 232 \\\hline
Pair 6& 0.47 & 76 & 36 \\\hline
Pair 7& 0.20 & 264 & 182\\\hline
Pair 8& 0.02 & 655 & 423 \\
\hline\end{tabular}
\caption{Statistics of Pairs vectors (binary) with their Similarity and Sparsity (non-zeros).}\vspace{-0.11in}
\label{tab:pairstat}
\end{table}

\begin{figure*}[t]
\begin{center}

\mbox{\hspace{-0.15in}
\includegraphics[width = 1.8in]{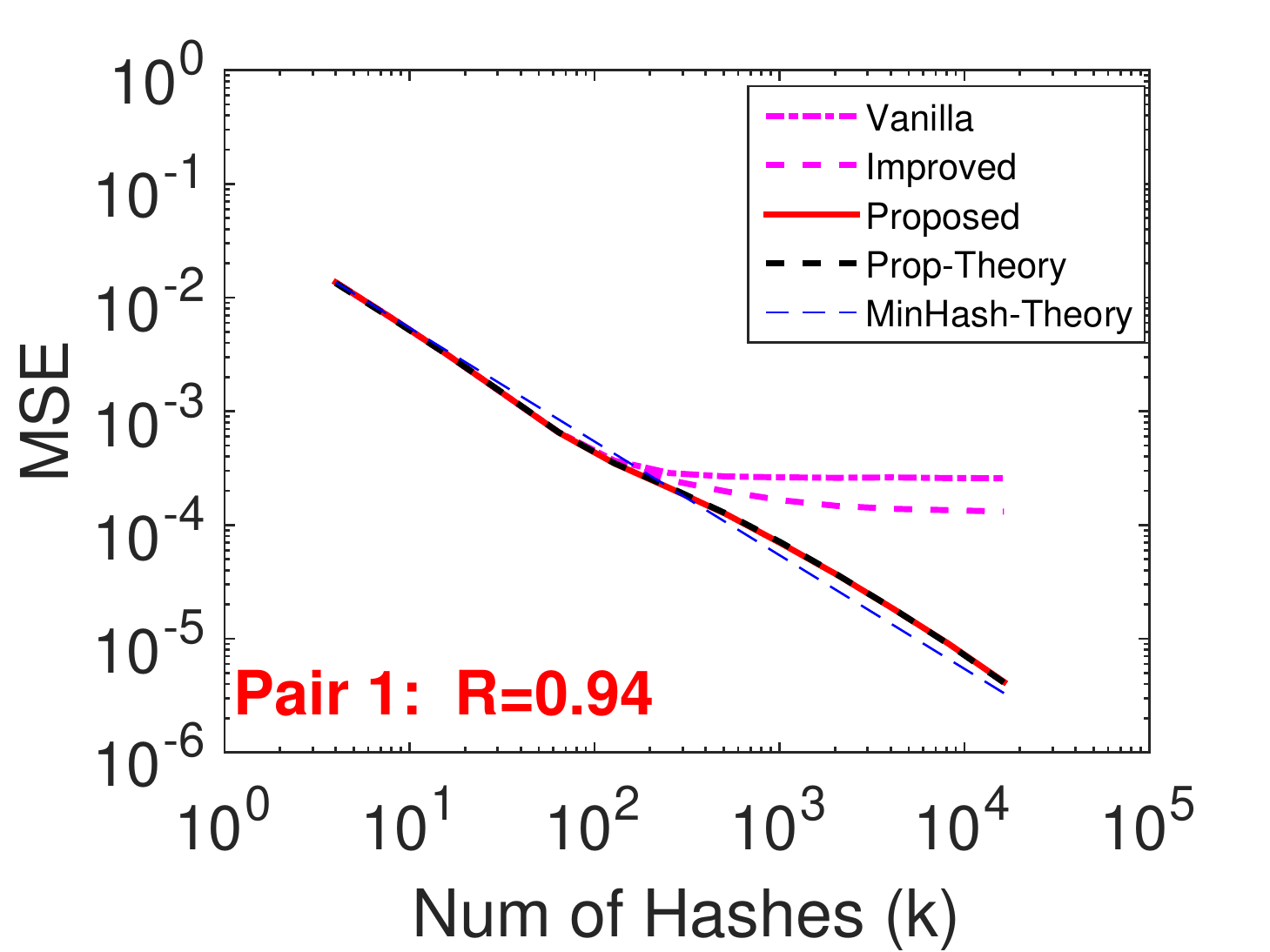}\hspace{-0.13in}
\includegraphics[width = 1.8in]{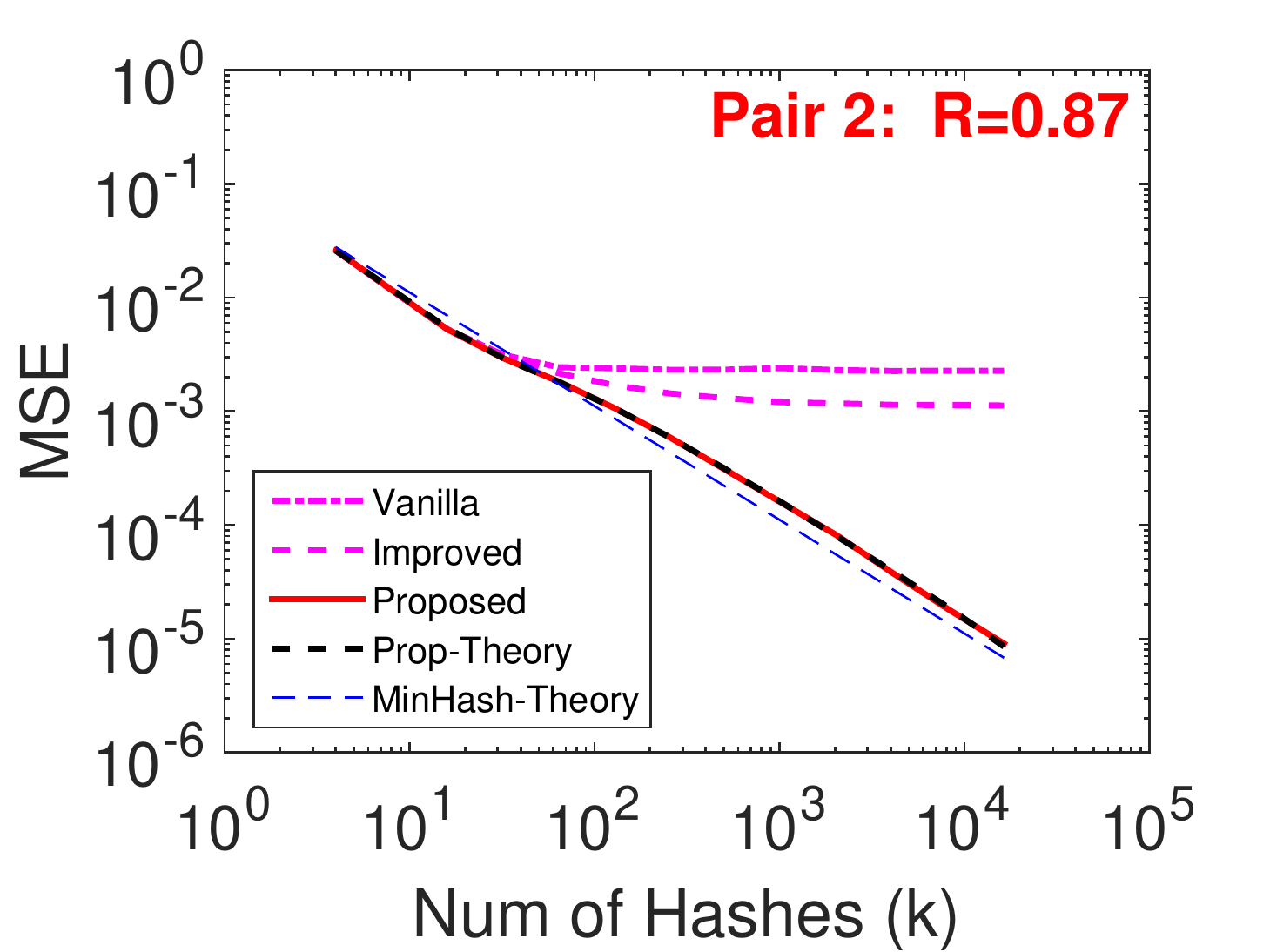}\hspace{-0.13in}
\includegraphics[width = 1.8in]{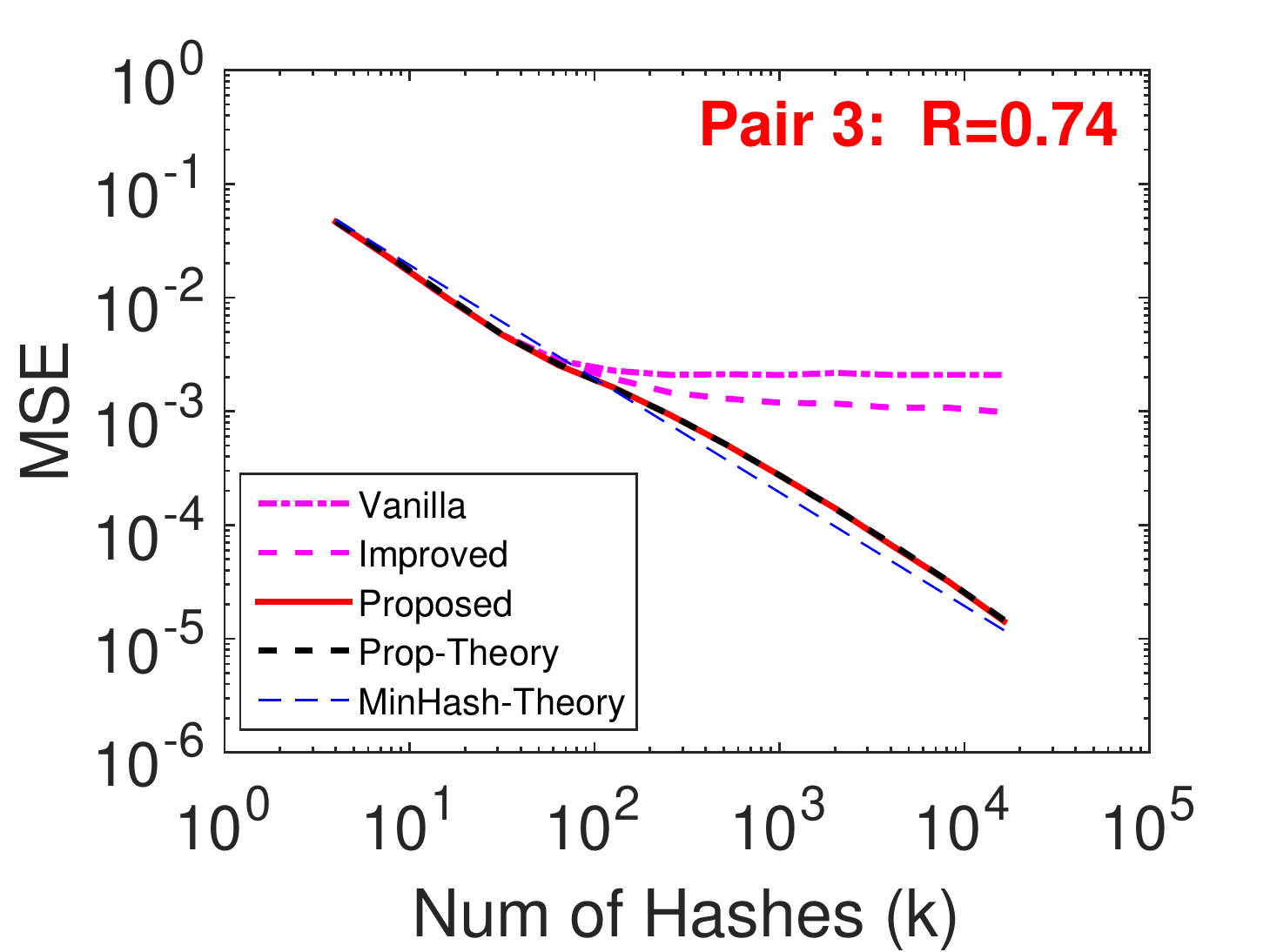}\hspace{-0.13in}
\includegraphics[width = 1.8in]{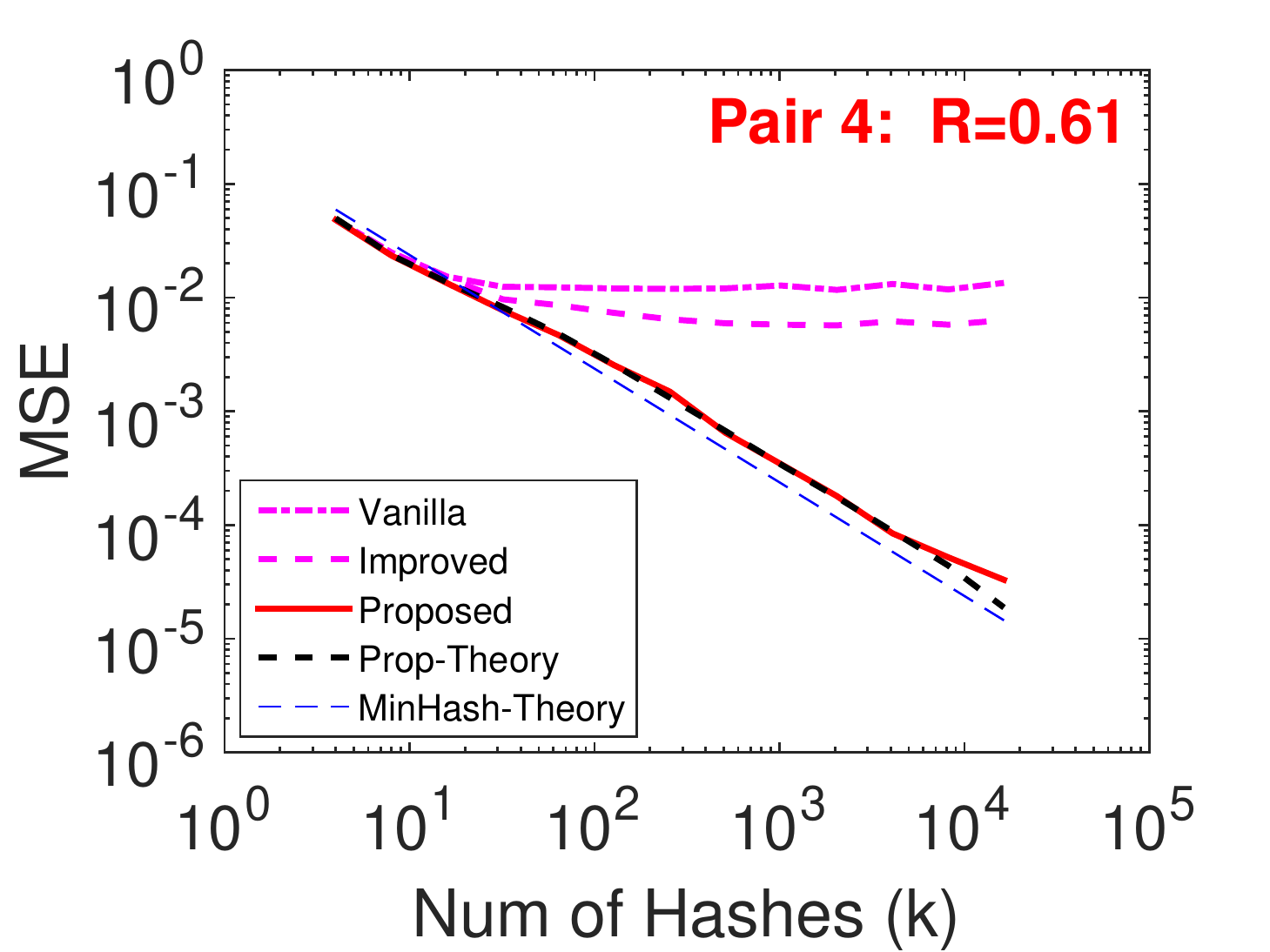}
}
\mbox{\hspace{-0.15in}
\includegraphics[width = 1.8in]{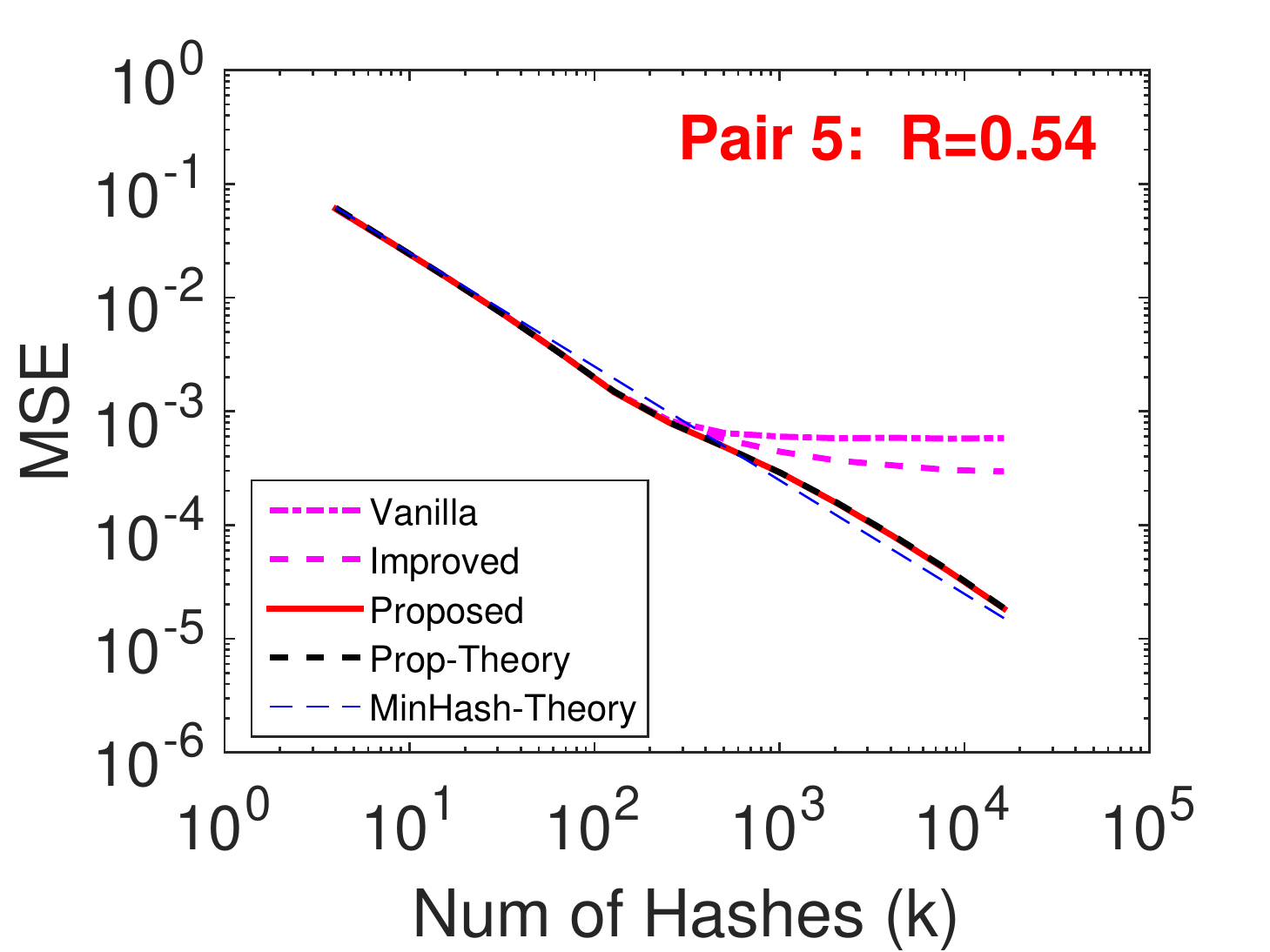}\hspace{-0.13in}
\includegraphics[width = 1.8in]{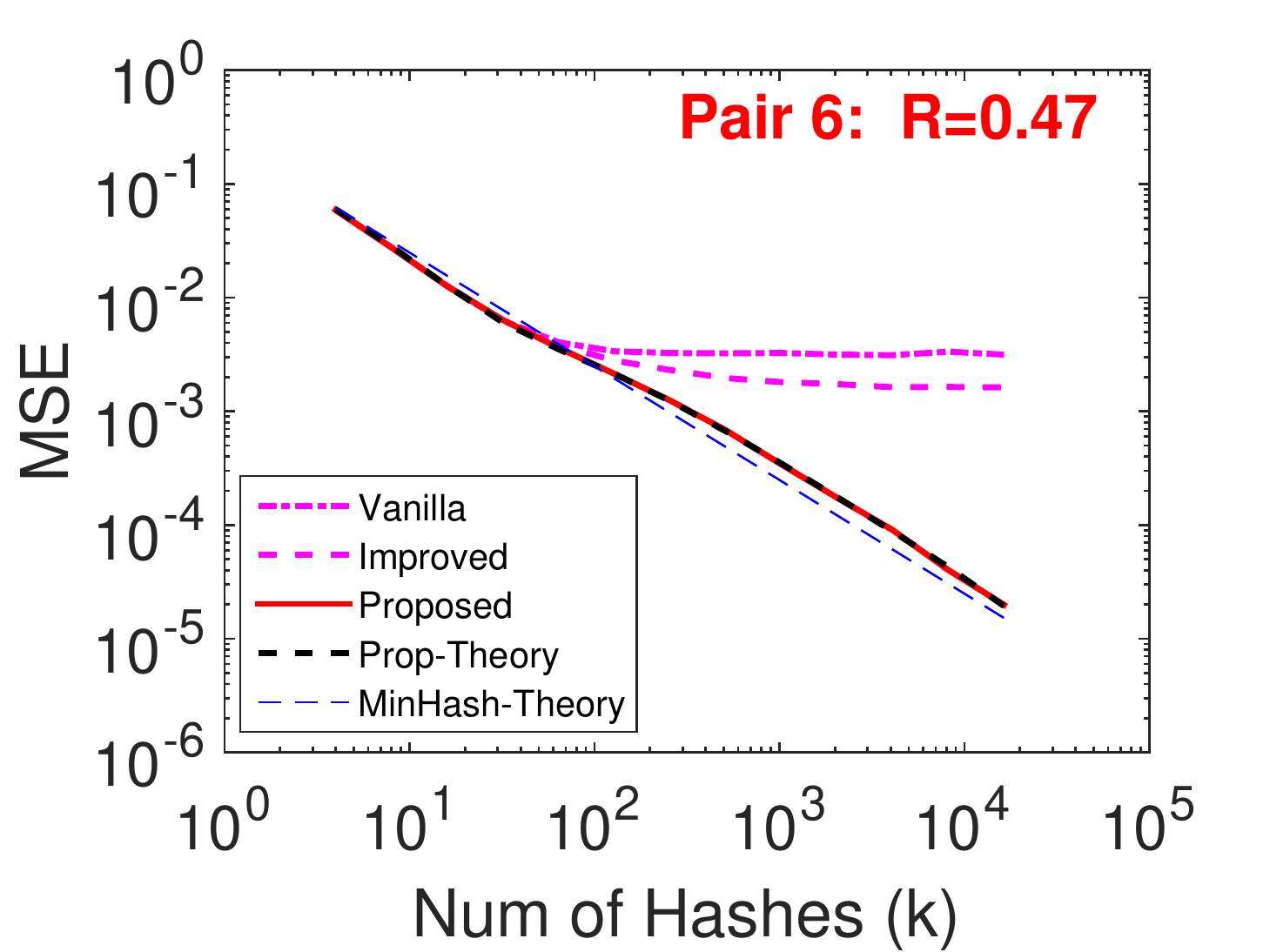}\hspace{-0.13in}
\includegraphics[width = 1.8in]{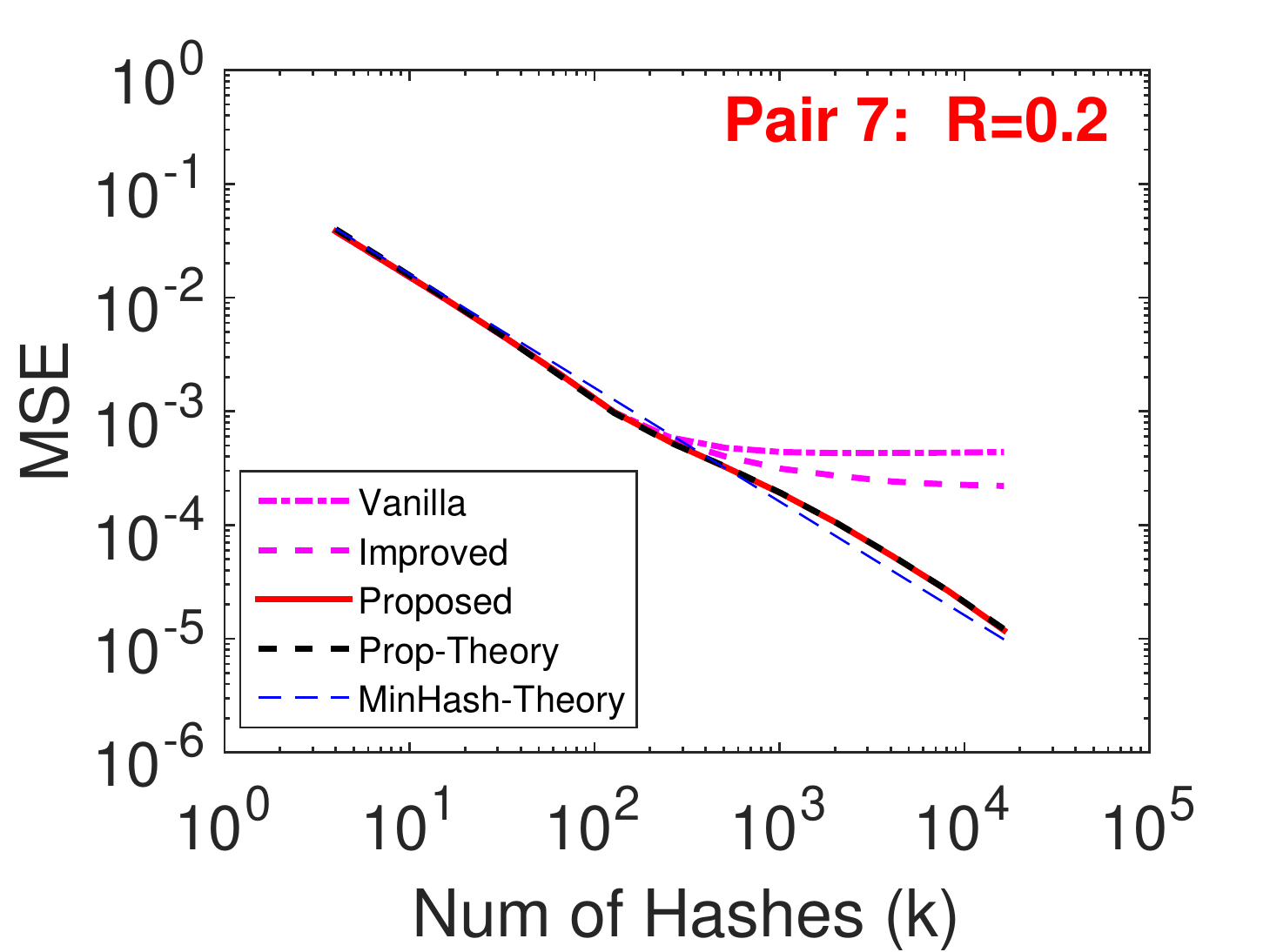}\hspace{-0.13in}
\includegraphics[width = 1.8in]{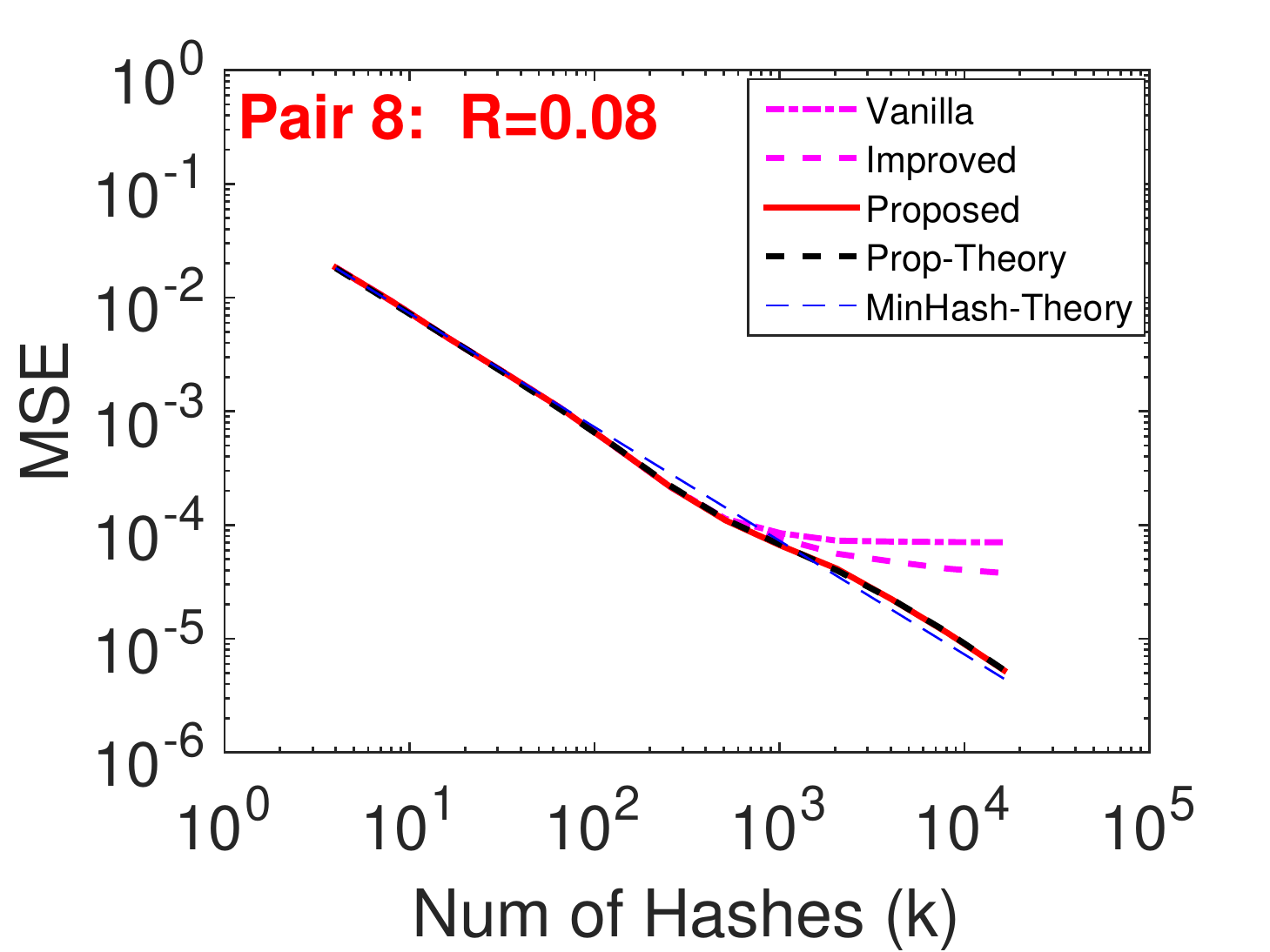}
}
\end{center}
\vspace{-0.15in}
\caption{\textbf{Average MSE} in Jaccard Similarity Estimation with the Number of Hash Values ($k$). Estimates are averaged over 5000 repetitions. With Optimal densification the variance is very close to costly minwise hashing, which is significantly superior to existing densification schemes. Note the log scale of y-axis.}
\label{fig:SimEst}\vspace{-0.1in}
\end{figure*}

We show that the expected running time of our proposal, including all constants, is very similar to the running time of the existing densification schemes.

\begin{table*}[ht]
\centering
  \begin{tabular}{|l|l|l|l||l|l|l||l|l|l|}
    \hline
    \multirow{2}{*}{} &
      \multicolumn{3}{|c||}{Existing $h^+$} &
      \multicolumn{3}{|c||}{This Paper $h^*$} &
      \multicolumn{3}{|c|}{Vanilla MinHash} \\\hline
    & 100 & 200 & 300 & 100 & 200 & 300 &100 &200 &300 \\
    \hline
    RCV1 & 62 & 86 & 104 & 56 & 90 & 116 & 376 & 733 & 1098 \\
    \hline
    URL & 373 & 584 & 624 & 287 & 459 & 631 & 2760 & 6247 & 8158\\
    \hline
    NEWS20 & 154 & 179 & 205 & 139 & 158 & 174 & 1201 & 2404 & 3650\\
    \hline
  \end{tabular}
  \caption{Time (in milliseconds) requires to compute 100, 200 and 300 hashes of the full data using the existing densification, proposed densification and vanilla minwise hashing. Minhash can be 10-18x slower for computing 300 hashes on these datasets.} \vspace{-0.11in}
  \label{tab:runningtime}
\end{table*}

Given set $S$ with $|S| =d$, we are interested in computing $k$ hash values.
The first step involves computing $k$ one permutation hashes (or sketches)
which only requires a single pass over the elements of $S$. This takes
$\max\{d,k\} \le d + k$ time. Now the densification
Algorithm~\ref{alg:optdens} requires a for loop of size $k$ and within each
for loop, if the bin is empty, it requires an additional while loop. Let
$N_{emp}$ be the number of empty bins, and therefore $\frac{k - N_{emp}}{k}$
is the probability that the while loop will terminate in one iteration (next
is not empty). Therefore, the expected number of iteration that each while
loop will run is a binomial random variable with expectation $\frac{k}{k
-N_{emp}}$. Thus, the expected running time of the algorithm is given by
\begin{align}\notag
  E[\mbox{Running Time}] &= d + 2k - N_{emp} + N_{emp}\frac{k}{k - N_{emp}}\\\notag
   &\le d + (r+2)k \ \mbox{where } \ r = \frac{N_{emp}}{N_{not-emp}}
\end{align}
The quantity $r$, which is the ratio of number of empty bins to the number of non-empty bins, is generally very small. It is rarely more than 2 to 3 in practice. Observe that randomly throwing $d$ items into $k$ bins, the expected number of empty bins is $E[N_{emp}] \approx k(1-\frac{1}{k})^d \approx ke^{-\frac{d}{k}}$. Which makes $r \approx \frac{e^{-d/k}}{1 - e^{-d/k}}$. The number of sketches is usually of the order of non-zeros. Even for very good concentration, the size of the sketches $k$ is rarely much larger than the size of the set $d$. Even when $k$ is 4 times $d$ the value of $r$ is approximately $3.5$. Thus, the quantity $r$ is negligible.

It should be further noted that the implementation cost of densification scheme $h^+$ is $d + 4k$ which in not very different from the cost of our proposal.
\section{Evaluations}

Our aim is to verify the theoretical claims of these papers empirically. We show that our proposal can replace minwise hashing for all practical purposes. To establish that, we focus on experiments with the following objectives:
\begin{enumerate}\vspace{-0.1in}
  \item Verify that our proposed scheme has a significantly better accuracy (variance) than the existing densification schemes. Validate our variance formulas.\vspace{-0.1in}
  \item Empirically quantify the impact of optimal variance in practice. How does this quantification change with similarity and sparsity? Verify that the proposal has accuracy close to vanilla minwise hashing.\vspace{-0.1in}
  \item Verify that there is no impact on running time of the proposed scheme over existing densification schemes, and our proposal is significantly faster than vanilla minwise hashing. Understand how the running time changes with change in sparsity and $k$?\vspace{-0.1in}
\end{enumerate}

\begin{table}[h]
\centering
\begin{tabular}{|c|c|c|c|} \hline
    & Avg. non-zeros (d) & Dim (D) & Samples \\\hline
RCV1& 73 & 47,236 & 20,242 \\\hline
URL & 115 & 3,231,961 & 100,000 \\\hline
News20 & 402 & 1,355,191 & 19,996 \\
\hline\end{tabular}
\caption{Basic Statistics of Datasets.}\vspace{-0.31in}
\label{tab:stat}
\end{table}

\subsection{Accuracy}

For objectives 1 and 2, we selected 9 different word pairs embedding, generated from new20 corpus, with varying level of similarity and sparsity. We use the popular term-document vector representation for each word.   The statistics of these word vector pairs are summarized in Table~\ref{tab:pairstat}

For each word pairs, we generated $k$ hashes using three different schemes: 1) Densification $h$, 2) Improved Densification $h^+$ and the proposed densification $h^*$ (Algorithm~\ref{alg:optdens}). Using these hashes, we estimate the Jaccard similarity (Equation~\ref{eq:est}). We plot the mean square error (MSE) with varying the number of hashes. Since the process is randomized, we repeat the process 5000 times, for every $k$, and report the average over independent runs. We report all integer values of $k$ in the interval $[1,2^{14}]$.

It should be noted that since all three schemes have the {\em LSH Property}, the bias is zero and hence the MSE is the theoretical variance. To validate our variance formula, we also compute and plot the theoretical value of the variance (Equation~\ref{eq:thevar}) of the optimal scheme. Also, to understand how all these fast methodologies compare with the accuracy of vanilla minwise hashing we also plot the theoretical variance of minwise hashing which is $\frac{R(1-R)}{k}$.

From the results in Figure~\ref{fig:SimEst}, we can conclude.\\
{\bf Conclusion 1:} The proposed densification is significantly more accurate, irrespective of the choice of sparsity and similarity, than the existing densification schemes especially for large $k$. Note the y-axis of plots is on log scale, so the accuracy gains are drastic.\\
{\bf Conclusion 2:} The gains with optimal densification is more for sparse data. \\
{\bf Conclusion 3:} The accuracy of optimal densification is very close the accuracy of costly minwise hashing.\\
{\bf Conclusion 4:} The theoretical variance of our proposal overlaps with the empirical estimates, and it seems to go to zero validating Theorem~\ref{eq:thevar}.\\
{\bf Conclusion 5:} The variances (or the MSE) of existing densification seems to converge to constant and do not go to zero confirming Theorem~\ref{theo:VarConst}.

\subsection{Speed}

\begin{table}[h]
\centering
\vspace{-0.11in}
\begin{tabular}{|c|c|c|c|} \hline
    & 100 Bins & 200 Bins & 300 Bins \\\hline
RCV1& 53 & 143 & 238 \\\hline
URL & 33 & 112 & 202 \\\hline
News20 & 14 & 58 & 120 \\
\hline\end{tabular}
\caption{Avg. Number of Empty Bins per vector (rounded) generated with One Permutation Hashing~\cite{Proc:Li_Owen_Zhang_NIPS12}. For sparse datasets, a significant (80\% with 300 hashes) of the bins can be empty and have no information. They cannot be used for indexing and kernel learning. Fortunately, we can efficiently densify them with optimal densification, such that, all the generated $k$ hashes have variance similar to minwise hashing. The overall computational cost is significantly less compared to minwise hashing}\vspace{-0.21in}
\label{tab:statebin}
\end{table}

To compute the runtime, we use three publicly available text datasets: 1) RCV1, 2) URL and 3) News20. The dimensionality and sparsity of these datasets are an excellent representative of the scale and the size frequently encountered in large-scale data processing systems, such as Google's SIBYL~\cite{Report:Sibyl}. The statistics of these datasets are summarized in Table~\ref{tab:stat}.

We implemented three methodologies for computing hashes: 1) Densification Scheme $h^+$, 2) The Proposed $h^*$ (Algorithm~\ref{alg:optdens} and 3) Vanilla Minwise Hashing. The methods were implemented in C++. Cheap hash function replaced costly permutations. Clever alternatives to avoid mod operations were employed. These tricks ensured that our implementations\footnote{Codes are available at \url{http://rush.rice.edu/fastest-minwise.html}} are as efficient as the possible. We compute the wall clock time required to calculate 100, 200 and 300 hashes of all the three datasets. The time include the end-to-end hash computation of the complete data. Data loading time is not included. The results are presented in Table~\ref{tab:runningtime}. All the experiments were done on Intel i7-6500U processor laptop with 16GB RAM.

Also, to get an estimate of the importance of densification, we also show the average number of empty bins generated by only using one permutation hashing and report the numbers in Table~\ref{tab:statebin}. We can clearly see that the number of empty bins is significantly larger and the hashes are unusable without densification.

From the running time numbers Table~\ref{tab:runningtime}, we conclude:

{\bf Conclusion 1:} Optimal densification is as fast as traditional densification irrespective of $k$ and the sparsity. However, optimal densification is significantly more accurate.\\
{\bf Conclusion 2:} Both the densification scheme is significantly faster than minwise hashing. They are 10-18x faster for computing 300 hashes on the selected datasets.

Given the simplicity, we hope our work gets adopted.

\bibliography{mybib_merged}
\bibliographystyle{icml2016}


\appendix

\section{Proofs}

\begin{theorem}
\label{theo:VarConst}
  Give any two finite sets $S_1, S_2 \in \Omega$, with $A = |S_1 \cup S_2|  > a = |S_1 \cap S_2| > 0$ and $|\Omega| = D \rightarrow \infty$. The limiting variance of the estimators from densification and improved densification when $k =D \rightarrow \infty$ is given by:
  \begin{small}
  \begin{align}\label{eq:limit1}
    \lim_{k \rightarrow \infty} Var(h) =  \frac{a}{A}\bigg[\frac{A -a}{A(A+1)}\bigg] &> 0  \\\label{eq:limit2}
    \lim_{k \rightarrow \infty} Var(h^+) =  \frac{a}{A}\bigg[\frac{3(A-1) + (2A-1)(a-1)}{2(A+1)(A-1)}& - \frac{a}{A}\bigg]> 0
  \end{align}
  \end{small}
\end{theorem}
{\bf Proof:} When $k = D$, then  $N_{emp} = D- A$. Substituting this value in the variance formulas from~\cite{Proc:Shrivastava_UAI14} and taking the limit as $D=k \rightarrow \infty$, we get the above expression after manipulation. When $0< R = \frac{a}{A} < 1$, they both are strictly positive. $\hfil\square$

\begin{theorem}
\label{theo:optDens}
  \begin{align}\label{eq:optDens}
    Pr\big(h^*(S_1) &= h^*(S_2)\big) = \frac{|S_1 \cap S_2|}{|S_1 \cap S_2|}= R \\\label{eq:thevar}
    Var(h^*) &= \frac{R}{k} + A \frac{R}{k^2} + B  \frac{R\bar{R}}{k^2} - R^2 \\
    \lim_{k \rightarrow \infty} Var(h^*) &= 0
  \end{align}
  where $N_{emp}$ is the number of simultaneous empty bins between $S_1$ and $S_2$ and the quantities $A$ and $B$ are given by
  \begin{small}
  \begin{align}\notag
    A &= \mathbb{E}\bigg[2N_{emp} + \frac{N_{emp}(N_{emp}-1)}{k-N_{emp}}\bigg] \\\notag
    B &=  \mathbb{E}\bigg[(k-N_{emp})(k - N_{emp}-1) + 2N_{emp}(k-N_{emp}-1)\\\notag &+ \frac{N_{emp}(N_{emp}-1)(k-N_{emp}-1)}{k-N_{emp}}\bigg]
  \end{align}
  \end{small}
\end{theorem}
{\bf Proof:}

The collision probability is easy using a simple observation that values coming from different bin numbers can never match across $S_1$ and $S_2$, i.e. $h^*_i(S_i) \ne h^*_j(S_2)$ if $i \ne j$, as they have disjoint different range. So whenever, for a simultaneous empty bin $i$, i.e. $E_i =1$, we get  $h^*_i(S_1) = h^*_i(S_2)$ after reassignment, the value must be coming from same non-empty bin, say numbers $k$ which is not not empty.
Thus,
$$Pr(h^*_i(S_1) = h^*_i(S_2)) = Pr(h^*_k(S_1) = h^*_k(S_2)| E_k =0)= R$$

The variance is little involved. From the collision probability, we have the following is unbiased estimator.
\begin{equation}
\hat{R} = \frac{1}{k}\sum_{j=0}^{k-1} \mathbbm{1}\{h^*_{j}(S_1) = h^*_{j}(S_2)\}.
\end{equation}
For variance, define the number of simultaneously empty bins by
\begin{align}
N_{emp} = \sum_{j =0}^{k-1}\mathbbm{1}\{E_j =1\},
\end{align}
where $\mathbbm{1}$ is the indicator function. We  partition the event $\left(h^*_j(S_1) = h^*_j(S_2)\right)$ into two cases depending on $E_j$. Let $M_j^N$ ({\bf Non-empty Match at $j$}) and $M_j^E$ ({\bf Empty Match at $j$}) be the events defined as:
\begin{align}
M_j^N &= \mathbbm{1}\{E_j =0\ \mbox{ and }\ h^*_j(S_1) =h^*_j(S_2)\} \\
M_j^E &= \mathbbm{1}\{E_j =1\ \mbox{ and }\ h^*_j(S_1) = h^*_j(S_2)\}
\end{align}
Note that,
$M_j^N = 1 \implies M_j^E = 0$ and $M_j^E = 1 \implies M_j^N = 0.$
From the  LSH property of estimator we have
\begin{align}
\mathbb{E}(M_j^N| E_j = 0) &= \mathbb{E}(M_j^E | E_j =1) \notag\\
&=\mathbb{E}( M_j^E + M_j^N ) = R  \ \ \forall j
\end{align}
It is not difficult to show that,
\begin{align}\notag
\mathbb{E}\left(M_j^NM_i^N\big| i \ne j, E_j = 0 \ \text{ and}  \ E_i = 0\right) = R\tilde{R},
\end{align}
where $\tilde{R} = \frac{a-1}{f1 + f2 - a -1}$.  Using these new events, we have
\begin{align}
\hat{R} = \frac{1}{k}\sum_{j=0}^{k-1} \left[ M_j^E + M_j^N \right]
\end{align}
We are interested in computing
\begin{align}
Var(\hat{R}) = \mathbb{E}\left(\left(\frac{1}{k}\sum_{j=0}^{k-1} \left[ M_j^E + M_j^N \right]\right )^2\right) - R^2
\end{align}

For notational convenience we will use $m$ to denote the event $k - N_{emp}=m$, i.e., the expression $\mathbb{E}( . | m)$ means $\mathbb{E}( . |k - N_{emp} = m).$ To simplify the analysis, we will first compute the conditional expectation
\begin{align}
f(m) = \mathbb{E}\left(\left(\frac{1}{k}\sum_{j=0}^{k-1} \left[ M_j^E + M_j^N \right]\right )^2\bigg| \ m\right)
\end{align}
 By expansion  and  linearity of expectation, we obtain
\begin{align}
k^2f(m) =  \mathbb{E}\left[\sum_{i \ne j}M_i^N M_j^N\bigg|  m\right] + \mathbb{E}\left[\sum_{i \ne j}M_i^N M_j^E\bigg|m\right]\notag \\
    + \mathbb{E}\left[\sum_{i \ne j}M^E_i M^E_j\bigg|m\right]  + \mathbb{E}\left[\sum_{i =1}^{k}\left[(M_j^N)^2 +(M_j^E)^2\right]\bigg| m\right]\notag
\end{align}
$M_j^N = (M_j^N )^2 $ and $M_j^E = (M_j^E)^2$ as they are indicator functions and can only take values 0 and 1. Hence,
\begin{align}
\label{eq:exp_square}
 \mathbb{E}\left[\sum_{j =0}^{k-1}\left[(M_j^N)^2 +(M_j^E)^2\right]\bigg| m\right]= kR
\end{align}
The values of the first three terms are given by the following 3 expression using simple binomial enpension and using the fact that we are dealing with indicator random variable which can only take values 0 or 1.
\begin{align}
\mathbb{E}\left[\sum_{i \ne j}M_i^N M_j^N\bigg|  m\right]  =  m(m-1)R\tilde{R}
\end{align}
\begin{align}
\mathbb{E}\left[\sum_{i \ne j}M_i^N M_j^E\bigg|m\right]  =  2m(k-m)\left[\frac{R}{m} + \frac{(m-1)R\tilde{R}}{m}\right]
\end{align}

Let $p$ be the probability that two simultaneously empty bins $i$ and $j$ finally picks the same non-empty bin for reassignment. Then we have
\begin{align}
\label{eq}
\mathbb{E}\left[\sum_{i \ne j}M^E_i M^E_j\bigg|m\right]= (k-m)(k-m-1)\left[pR + (1- p)R\tilde{R}\right]
\end{align}
because with probability $(1-p)$, it uses estimators from different simultaneous non-empty bin and in that case the $M^E_i M^E_j = 1$ with probability $R\tilde{R}$. We know that Algorithm~\ref{alg:optdens} which uses 2-universal hashing the value of $p = \frac{1}{m}$. This is because any pairwise assignment is perfectly random with 2-universal hashing.

Substituting for all terms with value of $p$ and rearranging terms gives the required expression.

When $k = D$, then  $N_{emp} = D- A$. Substituting this value in the variance formulas and taking the limit as $D=k \rightarrow \infty$, we get 0 for all $R$.

\begin{theorem}
  \begin{align}\label{eq:varcomp}
      Var(h^*) \le Var(h^+) \le Var(h)
  \end{align}
\end{theorem}
{\bf Proof:} We have $p* = \frac{1}{m} \le p^+ = \frac{1.5}{m+1} \le p = \frac{2}{m+1}$. The value of $p^+$ and $p$ comes from analysis in~\cite{Proc:Shrivastava_UAI14}

\begin{theorem}
Among all densification schemes, where the reassignment process for bin $i$ is independent of the reassignment process of any other bin $j$, Algorithm~\ref{alg:optdens} achieves the best possible variance.\end{theorem}
Under any independent re-assigment, the probability that two empty bins chooses the same non-empty bin out of $m$ non-empty bins is lower bounded by $\frac{1}{m}$ which is achieved by optimal densification.

\end{document}